\documentclass[3p,10pt,a4paper,twoside,fleqn,sort&compress]{filomat}
\usepackage{amssymb,amsmath,latexsym}
\usepackage{graphicx}%
\usepackage{amsfonts}%
\usepackage{amsthm}%
\usepackage{float}
\usepackage{caption}
\usepackage[varg]{pxfonts}

\newtheorem{theorem}{Theorem}[section]

\newtheorem{definition}[theorem]{Definition}

\newcommand{\FilVol}{}
\newcommand{\FilYear}{}
\newcommand{\FilPages}{}
\newcommand{\FilDOI}{}
\newcommand{\FilReceived}{}

\begin{document}

%

\title{Maximum values of the Sombor–index–like graph invariants of trees and connected graphs}

\author[affil1]{Milan Ba\v si\'c \corref{mycorrespondingauthor}}
\ead{basic\_milan@yahoo.com}
\address[affil1]{Faculty of Sciences and Mathematics, University of Ni\v s}
\newcommand{\AuthorNames}{M. Ba\v si\'c}

\newcommand{\FilMSC}{05C09, 11A07, 26C15, 26C10.}
\newcommand{\FilKeywords}{ Vertex-degree-based invariants, Molecular trees, Extremal values.}
\newcommand{\FilCommunicated}{}
\cortext[mycorrespondingauthor]{}
\newcommand{\FilSupport}{}

\begin{abstract}
A set of novel vertex-degree-based invariants was introduced by Gutman, denoted by \newline $SO_1, SO_2, \ldots,SO_6$. These invariants were constructed through geometric reasoning based on a new graph invariant framework. Motivated by proposed open problems in [Z. Tang, Q. Li, H. Deng, \textit{Trees with Extremal Values of the Sombor–Index–Like Graph Invariants}, MATCH Commun. Math. Comput. Chem. \textbf{90} (2023) 203–222], we have found the maximum values of $SO_5$ and $SO_6$ in the set of molecular trees with a given number of vertices, respectively, and we have found the maximum value of $SO_5$ in a class of connected graphs.
\end{abstract}

\maketitle

\makeatletter
\renewcommand\@makefnmark%
{\mbox{\textsuperscript{\normalfont\@thefnmark)}}}
\makeatother

\section{Introduction}\label{sec1}

The vertex-degree-based (VDB) topological indices quickly gained widespread attention due to their potential applications in both mathematics \cite{relations, cruz_rada, filomat, symmetry, temo, open_sombor, horoldagva, song, rada, shang, tensor, oboudi} and chemistry \cite{polymers, dendrimer, random, boiling, redzepovic}. In the past few decades, the focus of interest has been on the algebraic and combinatorial aspects of VDB indices. 
Recently, Gutman proposed several VDB graph invariants  $SO_1$, $SO_2$, $SO_3$, $SO_4$, $SO_5$, $SO_6$, which can be constructed by utilizing geometric arguments \cite{Gutman}. Gutman pointed out at the end of the paper \cite{Gutman} that it would be
interesting to examine the properties of these geometry-based topological
indices and see if these are useful in applications. 
They demonstrate mathematical characteristics different than other degree-based indices like the Sombor index, Zagreb indices, Forgotten topological index, and Randi\'c index, making them interesting from a mathematical standpoint for further exploration.
  
Mathematical and chemical properties of these geometry-based invariants have been considered recently. 
In \cite{Hechao}, the authors determined certain types of trees and connected graphs that attain the maximum and minimal values of the Sombor index $SO$.
More recently, some of the bounds for $SO_1$, $SO_2$, $SO_3$, $SO_4$, $SO_5$, $SO_6$ (Sombor-index-like graph invariants) among the classes of connected graphs and
(molecular) trees, with fixed numbers of vertices are obtained, and those
molecular trees that attain the extremal values are characterized for $SO_1$, $SO_2$, $SO_3$ and $SO_4$ \cite{Tang}.
The usefulness of the geometry-based invariants in modeling the thermodynamic properties of alkanes was tested, such as the acentric factor, entropy, and enthalpy of vaporization, and they were found to be effective predictors. Additionally, the chemical applicability of Sombor-index-like graph invariants was investigated and demonstrated that almost all of these indices are more accurate predictors of physicochemical properties than some commonly used indices (such as the first Zagreb index ($M_1$), forgotten topological index
($F$), connectivity index ($R$), sum connectivity index ($SCI$) and the Sombor
index ($SO$))\cite{Tang}. Most recently, extremal values for invariants $SO_3$, $SO_4$, $SO_5$, and $SO_6$ have been obtained in the class of trees, as well as in the class of connected graphs for invariants $SO_3$ and $SO_4$. Furthermore, sharp bounds for invariants $SO_5$ and $SO_6$ have been established \cite{ali}.

Motivated by the proposed open problems in \cite{Tang}, the solutions to which would enhance the comprehensiveness of studying Sombor-index-like graph invariants, we have determined the maximum values of certain Sombor-index-like graph invariants and identified extremal graphs within the classes of molecular trees and a specific class of connected graphs with a fixed number of vertices. In particular, we determine the maximum values of $SO_5$ and $SO_6$, respectively, within the set of molecular trees with a specified number of vertices. Additionally, we ascertain the maximum value of $SO_5$ within the set of graphs resulting from the join operation applied to two specific graphs of a given order. We posit that the solution to the latter problem also addresses the general problem: finding the maximum value of $SO_5$ within the class of connected graphs and their corresponding extremal graphs.
Generally, the proofs presented in this paper are based on the connection between number theory, polynomial theory and  multivariable function analysis, and fall into a good many distinct cases. Attempts to determine the maximum value of $SO_6$ in the set of connected graphs 
would be much more demanding, likely involving the examination of a significantly greater number of cases and more complex mathematical techniques than those  presented in the preceding studies (such as inequality theory and elementary graph theory). 
A more detailed insight into unresolved aspects of the problem is provided through certain directions for future research outlined in the concluding remarks section.

\bigskip

Let $d_G(u)$ denote the degree of vertex $u$ in graph $G$, with $V(G)$ and $E(G)$ representing the sets of vertices and edges, respectively. Additionally, we define $\delta(G)$ and $\Delta(G)$ as the minimum and maximum degrees of graph $G$. These notations together with the following definition will serve as the main terminology throughout the paper.

\smallskip

\begin{definition}
For a connected graph $G$ with the set of edges $E(G)$, Sombor-index-like vertex-degree-based graph invariants, labeled as $SO_5$ and $SO_6$ are defined as

\begin{eqnarray*}
   SO_5(G) &=& 2\pi \sum_{uv \in E(G)}\frac{\mid d_G^2(u) - d_G^2(v) \mid}{\sqrt{2} + 2\sqrt{d_G^2(u) + d_G^2(v)}}\\
   SO_6(G) &=&\pi \sum_{uv \in E(G)}\left( \frac{d_G^2(u) - d_G^2(v)}{\sqrt{2} + 2\sqrt{d_G^2(u) + d_G^2(v)}} \right)^2.
\end{eqnarray*}
\end{definition}

Let the functions $f$ and $g$ be defined as $f(a, b) = \frac{\mid a^2 - b^2 \mid}{\sqrt{2}+2\sqrt{a^2+b^2}}$, and $g(a, b) = f^2(a,b)$.


\section{Maximum values of $SO_5$ and $SO_6$ in the class of molecular trees}

A molecular tree, denoted by $MT$, is a tree with a maximum degree not greater than $4$. In this section, we will establish the maximum values of the functions $SO_5$ and $SO_6$ in the set of connected molecular trees.

\begin{theorem}
\label{molekularna}
    Let $MT$ be a connected molecular tree of order $n$ ($n \geq 5$), then the following inequalities hold
    \begin{itemize}
        \item[(i)] $SO_5(MT) \leq \begin{cases} 
            2\pi\left(\frac{f(1,4)+f(2,4)}{2}n +\frac{3f(1,4)-5f(2,4)}{2}\right), &  n \equiv 1 \pmod 4\\
            2\pi\left( \frac{f(1,4)+f(2,4)}{2}n +f(1,2)-2f(2,4) \right), & n \equiv 2 \pmod 4\\
            2\pi\left( \frac{f(1,4)+f(2,4)}{2}n+\frac{2f(3,4)+4f(1,3)-f(1,4)-7f(2,4)}{2} \right), & n \equiv 3 \pmod 4\\
            2\pi\left( \frac{f(1,4)+f(2,4)}{2}n+2f(1,4)-4f(2,4) \right), & n \equiv 0 \pmod 4
        \end{cases}$
        \item[(ii)] $SO_6(MT) \leq \begin{cases}
            \pi\left( \frac{g(1,4)+g(2,4)}{2}n +\frac{3g(1,4)-5g(2,4)}{2}  \right), & n \equiv 1 \pmod 4\\
            \pi\left( \frac{g(1,4)+g(2,4)}{2}n+2g(1,4)-5f(2,4)+g(3,4) \right), & n \equiv 2 \pmod 4\\
            \pi\left( \frac{g(1,4)+g(2,4)}{2}n +\frac{5g(1,4)-11g(2,4)}{2}  \right), & n \equiv 3 \pmod 4\\
            \pi\left( \frac{g(1,4)+g(2,4)}{2}n+2g(1,4)-4f(2,4) \right), & n \equiv 0 \pmod 4.\\
        \end{cases}$
    \end{itemize}

    \end{theorem}
\proof
Let $n_i$ be the number of vertices of the molecular tree with the degree equal to $i$, for $1 \leq i \leq 4$. The number of edges whose end-vertices of molecular trees are of the degrees $i$ and $j$, for $1 \leq i, j \leq 4$, are denoted as $m_{ij}$. The following equations hold for any graph with maximal degree $4$ of order $n$ and $m$ edges

\begin{align}\label{sistem}
\begin{split}
    &n_1 + n_2 + n_3 + n_4 = n,\\
    &n_1 + 2n_2 + 3n_3 + 4n_4 = 2m,\\
    &m_{12} + m_{13} + m_{14} = n_1,\\
    &m_{12} + 2m_{22} + m_{23} + m_{24} = 2n_2,\\
    &m_{13} + m_{23} + 2m_{33} + m_{34} = 3n_3,\\
    &m_{14} + m_{24} + m_{34} + 2m_{44} = 4n_4.\\
\end{split}
\end{align}

Using the fact that $m=n-1$, we observe that we can express the variables $n_1$, $n_2$, $n_3$, $n_4$ $m_{14}$, and $m_{24}$ in terms of $m_{12}$, $m_{13}$, $m_{22}$, $m_{23}$ and $m_{33}$.

The functions $SO_5(MT)$ and $SO_6(MT)$ can be expressed as $SO_5(MT) = 2\pi\displaystyle\sum_{ij \in E(MT)}f(d_i, d_j)$ and $SO_6(MT) = \pi\displaystyle\sum_{ij \in E(MT)}g(d_i, d_j)$. Employing the fact that $f(i, i) = g(i, i) = 0$, these functions can be written as follows

\begin{eqnarray}
\label{SO5}
    \frac{SO_5(MT)}{2\pi} &=&  \frac{n(f(1,4)+f(2,4))}{2}\cdot f(1,4) +\left(\frac{3f(1,4)}{2}-\frac{5f(2,4)}{2}\right)+ m_{12}\left(f(1,2) - \frac{3f(1,4)}{2}+\frac{f(2,4)}{2}\right)+ \nonumber\\  
    &+& m_{13}\left(f(1,3) - \frac{7f(1,4)}{6}+ \frac{f(2,4)}{6}\right) -m_{22}\left(\frac{f(1,4)}{2}+\frac{f(2,4)}{2}\right)+ \nonumber\\ 
    &+&m_{23}\left(f(2,3) - \frac{f(1,4)}{6}-\frac{5f(2,4)}{6}\right)+ m_{33}\left(\frac{f(1,4)}{6}-\frac{7f(2,4)}{6}\right) +\nonumber\\ &+&m_{34}\left(f(3,4)+\frac{f(1,4)}{3}-\frac{4f(2,4)}{3}\right)
    + m_{44}\left(\frac{f(1,4)}{2} - \frac{3f(2,4)}{2}\right) .\label{6}
\end{eqnarray}

Motivated by the fact that only non-negative constants are the ones multiplying $n$, we prove that molecular trees attaining maximum for $SO_5$ are trees whose adjacent vertices $u$ and $v$ satisfy that $(d_u, d_v) \in \{(1, 4), (2,4)\}$, if such tree exists. By analyzing modular arithmetic in the context of solving (\ref{sistem}), we can deduce that these trees exist for the order  $n\equiv 1 \pmod 4$.  
Considering the absolute values of negative constants that multiply $m_{ij}$ in (\ref{SO5}), further modular arithmetic analysis can be conducted to prove that trees achieving maximum values of $SO_5$ are those possessing only $(1,4)$ and $(2,4)$ edges, with precisely one $(1,2)$ edge if $n \equiv 2 \pmod 4$, one $(3,4)$ edge and two $(1,3)$ edges if $n \equiv 3 \pmod 4$, and precisely one $(4,4)$ edge if $n \equiv 0 \pmod 4$.

\medskip

In the same manner, we can demonstrate that $SO_6$ attains maximal values when considering trees containing only edges of type $(1,4)$ and $(2,4)$, specifically when $n \equiv 1 \pmod 4$. For the remaining cases, optimal trees include edges $(1,4)$ and $(2,4)$ along with one $(3,4)$ edge if $n \equiv 2 \pmod 4$, precisely two $(4,4)$ edges if $n \equiv 3 \pmod 4$, and exactly one $(4,4)$ edge if $n \equiv 0 \pmod 4$.

In all of the aforementioned cases, the specific number of $(1,4)$ and $(2,4)$ edges can be calculated using the system (\ref{sistem}).
\qed


\section{Maximum value of $SO_5$ in a class of connected graphs}

In this section we analyze which graphs are potential candidates that achieve maximum $SO_5$ in the class of all connected graphs of order $n$.

Suppose an extremal graph concerning the maximum value of $SO_5$ has only two values in its degree sequence, namely $\delta(G)=\delta$ and $\Delta(G)=\Delta$.
We notice that the vertices with degrees $\Delta$ constitute a clique of the size $l$ as a subgraph within $G$, where $l$ represents the number of vertices with degree $\Delta$, whereas vertices with degrees $\delta$ form an independent set.
If the converse were true, implying the absence of an edge between two vertices of degree $\Delta$, such as $u$ and $v$, upon introducing an edge between them, we would obtain a graph $G_1$. In this case, we have
$$
SO_5(G_1)-SO_5(G)=2\pi\sum_{uw\in E(G)} \left(f(\Delta+1,d_w)-f(\Delta,d_w)\right)+\sum_{vw\in E(G)} (f(\Delta+1,d_w)-f(\Delta,d_w)>0),
$$ 
which leads to a contradiction. Similarly, if an edge exists between two vertices of degree $\Delta$, such as $u$ and $v$, upon removing this edge, we would obtain a graph $G_2$ where
$$
SO_5(G_2)-SO_5(G)=2\pi\sum_{uw\in E(G)} (f(\delta,d_w)-f(\delta-1,d_w))+\sum_{vw\in E(G)} (f(\delta-1,d_w)-f(\delta,d_w)>0),
$$ 
which again contradicts our assumption. 
Hence, we deduce that the number of edges between vertices with degrees $\Delta$ and $\delta$ in graph $G$ equals $\delta(n-l)$, implying
$$
SO_5(G)= 2\pi\cdot\delta(n-l)f(\delta,\Delta)\leq 2\pi\cdot\delta(n-\delta)f(\delta,n-1) 
$$
since $\delta \leq l$ (all neighbors of a vertex with degree $\delta$ belong to vertices of degree $\Delta$).

Hence, we seek extremal graphs, denoted as $M_{n, k}$, among those with an order of $n$, constructed as the join of an empty graph of order $n-k$, denoted as $\overline{K_{n-k}}$, and a complete graph of order $k$, for $1\leq k\leq n$. 
This graph clearly contains vertices with degrees that are either $k$ or $n-1$, where $n_k = n-k$ and $n_{n-1} = k$.
Notice that 
\begin{eqnarray}
\label{eq1}
SO_5(M_{n,k}) = 2\pi\cdot k(n-k)f(k,n-1) = 2\pi \cdot k(n-k)\cdot\frac{(n-1)^2-k^2}{\sqrt{2}+2\sqrt{(n-1)^2 + k^2}}.
\end{eqnarray}

\bigskip

\begin{theorem}
    Let $G$ be a connected graph of the order $n (n>2)$ such that for every vertex $v\in V(G)$ holds that $d(v)\in \{\delta(G),\Delta(G)\}$, then the following inequality holds

    $$SO_5(G) \leq \underset{i \in \{\lfloor c_0n \rfloor - 1, \lfloor c_0n \rfloor , \lceil c_0n \rceil\}}{\max}SO_5(M_{n,i})$$

    where $c$ is given by the expression
    \begin{eqnarray*}
        c_0&=& -\frac{1}{12} \sqrt{4 \sqrt[3]{6 \sqrt{7422}+505}-\frac{92}{\sqrt[3]{6 \sqrt{7422}+505}}-31}-\frac{1}{12}+\\
        &+& \frac{1}{2}  \sqrt{\begin{aligned}-\frac{1}{9}& \sqrt[3]{6 \sqrt{7422}+505}-\frac{31}{18}+\frac{23}{9 \sqrt[3]{6 \sqrt{7422}+505}}+\\&+\frac{25}{18 \sqrt{4 \sqrt[3]{6 \sqrt{7422}+505}-\frac{92}{\sqrt[3]{6 \sqrt{7422}+505}}-31}}\end{aligned}}.
    \end{eqnarray*}


\end{theorem}

\proof


According to (\ref{eq1}), it holds that 
$$
SO_5(G) \leq 2\pi\cdot\delta(n-\delta)\cdot\frac{(n-1)^2-\delta^2}{\sqrt{2} + 2\sqrt{(n-1)^2 + \delta^2}} =2\pi\cdot F(\delta).
$$

Since $n>2$ is fixed, 
the right hand side of the inequality is a single-variate function, which will be analysed in the rest od the proof.
The function $F(\delta)$ attains maximum in some of the following cases: $\delta = 1$, or $\delta = n-1$, or $F'(\delta) = 0$. 

Case $\delta = 1$ is equivalent to $G$ being a star graph $S_n$. In this case, the value of the function is equal to $SO_5(S_n) =2\pi F(1) = 2\pi\frac{n^3-3n^2+2n}{\sqrt{2} + 2\sqrt{n^2-2n+2}}$.

The case where $\delta = n-1$ is equivalent to $G$ being a complete graph $K_n$. It is straightforward to calculate that $SO_5(K_n) = F(n-1) = 0$, which represents the minimum of $F$.

To find the maximum value of $F(\delta)$, it is necessary to determine its stationary points. It becomes evident that the first derivative of $F(\delta)$ is equal to
\begin{eqnarray*}
    F'(\delta) &=& \frac{((n-2\delta)((n-1)^2-\delta^2) + (n\delta-\delta^2)(-2\delta))(\sqrt{2}+2\sqrt{(n-1)^2+\delta^2})}{(\sqrt{2}+2\sqrt{(n-1)^2+\delta^2})^2} -\\
    &-& \frac{(n\delta-\delta^2)((n-1)^2-\delta^2)\cdot\frac{2\delta}{\sqrt{\delta^2+(n-1)^2}}}{(\sqrt{2}+2\sqrt{(n-1)^2+\delta^2})^2}\\
    &=&-\dfrac{2\left(n-\delta\right)\delta^2}{2\sqrt{\delta^2+\left(n-1\right)^2}+\sqrt{2}}-\dfrac{\delta\cdot\left(\left(n-1\right)^2-\delta^2\right)}{2\sqrt{\delta^2+\left(n-1\right)^2}+\sqrt{2}}+\\
    &+&\dfrac{\left(n-\delta\right)\left(\left(n-1\right)^2-\delta^2\right)}{2\sqrt{\delta^2+\left(n-1\right)^2}+\sqrt{2}}-\dfrac{2\left(n-\delta\right)\delta^2\cdot\left(\left(n-1\right)^2-\delta^2\right)}{\sqrt{\delta^2+\left(n-1\right)^2}\left(2\sqrt{\delta^2+\left(n-1\right)^2}+\sqrt{2}\right)^2}.\\
\end{eqnarray*}

After a brief calculation, the numerator of the first derivative is equal to 
\begin{eqnarray*}
    T(\delta) &=& 6 \delta^{5} - 4 \delta^{4} n + 6 \delta^{3} \left(n - 1\right)^{2} + \delta^{2} \left(- 6 n^{3} + 12 n^{2} - 6 n\right) +\\
    &+& \delta \left(- 4 n^{4} + 16 n^{3} - 24 n^{2} + 16 n - 4\right) + 2 n^{5} - 8 n^{4} + 12 n^{3} - 8 n^{2} + 2 n+\\
    &+& \sqrt{2 \delta^{2} + 2 \left(n - 1\right)^{2}} \cdot(4 \delta^{3} - 3 \delta^{2} n - 2 \delta \left(n - 1\right)^{2} + n \left(n - 1\right)^{2}).
\end{eqnarray*}
Our objective is to show that it possesses a unique zero within the interval $[1,n-1]$.
For some $c \in \mathbb{R}$, the function $T(cn)$ is equal to
\begin{eqnarray*}
    T(cn) &=& n^{5} \left(6 c^{5} - 4 c^{4} + 6 c^{3} - 6 c^{2} - 4 c + 2\right) + n^{4} \left(- 12 c^{3} + 12 c^{2} + 16 c - 8\right) +\\
    &+& n^{3} \left((4 c^{3} - 3c^2 - 2c + 1) \sqrt{n^{2} \left(2 c^{2} + 2\right) - 4 n + 2} + 6 c^{3} - 6 c^{2}  - 24 c + 12\right) +\\
    &+&n^{2} \left(4 c \sqrt{n^{2} \left(2 c^{2} + 2\right) - 4 n + 2} + 16 c - 2 \sqrt{n^{2} \left(2 c^{2} + 2\right) - 4 n + 2} - 8\right) +\\
    &+& n \left(- 2 c \sqrt{n^{2} \left(2 c^{2} + 2\right) - 4 n + 2} - 4 c + \sqrt{n^{2} \left(2 c^{2} + 2\right) - 4 n + 2} + 2\right).
\end{eqnarray*}

To determine the value of $c \in \mathbb{R}$ at which the leading coefficient vanishes, we need to find the roots of the polynomial $6c^5 - 4c^4 + 6c^3 - 6c^2 - 4c + 2 = 0$. It is observed that $c = 1$ is a root of this fifth-degree polynomial. Among the remaining four roots, two are complex numbers, one is negative, and the other is a positive root denoted as $c_0$, approximately equal to $0.365046124400441$ (the exact value is specified in the assertion of the theorem).


We prove that a zero of $T(\delta)$ belonging to the interval $[1, n-1]$ actually falls within the interval $(c_0n-0.5, c_0n)$. In the remainder of the proof, we will employ numerical notation instead of exact forms. This choice enhances readability, as in the most of cases we solely require the signs of the values, not their precise forms. 

We can calculate the values of the functions $T(c_0 n)$ and $T(c_0 n -0.5)$

\begin{eqnarray*}
    T(c_0 n) &=& - 1.144 n^{4} + n^{3} \Biggl(0.129 \sqrt{0.567 n^{2}  - n + \frac{1}{2}} + 2.731\Biggr)+\\ 
    &+&n^{2} \Biggl(- 1.08\cdot  \sqrt{0.567 n^{2} - n + \frac{1}{2}}- 2.16\Biggr) +  n \Biggl(0.54 \sqrt{0.567 n^{2} - n + \frac{1}{2}} + 0.54\Biggr)\\
    T(c_0 n - 0.5) &=& 1.97 n^{4} + n^{3} \Bigl(- 7.178+0.099 \cdot \sqrt{2}\sqrt{0.479 n^{2} - n + 0.528} \Bigr) +\\
    &+& n^{2} \Bigl( 9.527+1.162 \sqrt{2}\cdot\sqrt{0.479 n^{2} - n + 0.528}\Bigr) + \\
    &+& n \Bigl(-5.383 - 2.129 \sqrt{2}\cdot \sqrt{0.479 n^{2} - n + 0.528}\Bigr) + \\
    &+& 0.768 \sqrt{2}\sqrt{0.479 n^{2} - n + 0.528}+ 1.063.
\end{eqnarray*}

Now, we prove that $T(c_0n) < 0$, for $n \geq 3$. It is clear that the factor multiplying $n^4$ is smaller than $-1.14$, and the factor multiplying $n^3$ is smaller than $0.14n+2.74$. On the other hand, it holds that the factor multiplying $n^2$ is smaller than $-2$, and the factor multiplying $n$ is smaller than $n + 0.54$. This means that $T(c_0n) < -1.14n^4 + n^3(0.14n+2.74) - 2n^2 + n(n + 0.54)$. Notice that for $n \geq 3$ it holds that $-1.14n^4 + 0.14n^4 + 2.74n^3 < 0$ and $-2n^2 + n^2 + 0.54n < 0$, which implies that $T(c_0n) < 0$. 

We also prove that $T(c_0n-0.5) > 0$, for $n \geq 3$. It is clear that the factor multiplying $n^4$ is greater than $1.96$ and the factor multiplying $n^3$ is greater than $-7.2$. On the other hand, it holds that the factor multiplying $n^2$ is greater than $9.5$ and the factor multiplying $n$ is greater than $-2.13\sqrt{2}n-5.4$. Since the remaining summands are clearly positive, we can conclude that $T(c_0n-0.5) > 1.96n^4 - 7.2n^3 + 9.5n^2 -2.13\sqrt{2}n^2-5.4n$. For $n \geq 4$, it holds that $1.96n^4-7.2n^3 > 0$ and $9.5n^2-2.13\sqrt{2}n^2-5.4n > 0$. Through  direct computation, it can be calculated that $T(c_0n-0.5) > 0$ for $n = 3$. This implies that $T(c_0n-0.5) > 0$ for $n \geq 3$.

Since $T$ is continuous, we confirm the presence of a zero of $T(\delta)$ within the interval $(c_0n-0.5, c_0n)$, and in the subsequent part of the proof, we show that all roots exist within this interval on the domain $[1,n-1]$.
\bigskip

Now define two polynomials as follows $T_l(\delta)=T(\delta)-(\sqrt{2 \delta^{2} - 2 \left(n - 1\right)^{2}}+\sqrt{0^2+(n-1)^2} ) \cdot(4 \delta^{3} - 3 \delta^{2} n - 2 \delta \left(n - 1\right)^{2} + n \left(n - 1\right)^{2})$ and $T_u(\delta)=T(\delta)-(\sqrt{2 \delta^{2} - 2 \left(n - 1\right)^{2}}+\sqrt{2(n-1)^2} ) \cdot(4 \delta^{3} - 3 \delta^{2} n - 2 \delta \left(n - 1\right)^{2} + n \left(n - 1\right)^{2})$.
Since $\sqrt{0^2+(n-1)^2} \leq  \sqrt{\delta^2+(n-1)^2} \leq \sqrt{(n-1)^2+(n-1)^2}$, and $4 \delta^{3} - 3 \delta^{2} n - 2 \delta \left(n - 1\right)^{2} + n \left(n - 1\right)^{2} \geq 0$, for $n > 2$, it is clear that for $2 \leq \delta \leq n-1$, it holds that $T_l(\delta) \leq T(\delta) \leq T_u(\delta)$.
The precise forms of the polynomials $T_l(\delta)$ and $T_u(\delta)$ are presented below
\begin{eqnarray*}
    T_l(\delta) &=& 6 \delta^{5} - 4 \delta^{4} n + 6 \delta^{3} \left(n - 1\right)^{2} + \delta^{2} \left(- 6 n^{3} + 12 n^{2} - 6 n\right) +\\
    &+& \delta \left(- 4 n^{4} + 16 n^{3} - 24 n^{2} + 16 n - 4\right) + 2 n^{5} - 8 n^{4} + 12 n^{3} - 8 n^{2} + 2 n+\\
    &+& \sqrt{2\cdot 0^{2} + 2 \left(n - 1\right)^{2}} \cdot(4 \delta^{3} - 3 \delta^{2} n - 2 \delta \left(n - 1\right)^{2} + n \left(n - 1\right)^{2})\\
    &=&6 \delta^{5} - 4 \delta^{4} n + \delta^{3} \left(6 n^{2} - 12 n + 4 \sqrt{2} n - 4 \sqrt{2} + 6\right) +\\
    &+& \delta^{2} \left(- 6 n^{3} - 3 \sqrt{2} n^{2} + 12 n^{2} - 6 n + 3 \sqrt{2} n\right) +\\
    &+& \delta \left(- 4 n^{4} - 2 \sqrt{2} n^{3} + 16 n^{3} - 24 n^{2} + 6 \sqrt{2} n^{2} - 6 \sqrt{2} n + 16 n - 4 + 2 \sqrt{2}\right) +\\
    &+&2 n^{5} - 8 n^{4} + \sqrt{2} n^{4} -3 \sqrt{2} n^{3} + 12 n^{3} - 8 n^{2} + 3 \sqrt{2} n^{2} - \sqrt{2} n + 2 n\\
    T_u(\delta) &=& 6 \delta^{5} - 4 \delta^{4} n + 6 \delta^{3} \left(n - 1\right)^{2} + \delta^{2} \left(- 6 n^{3} + 12 n^{2} - 6 n\right) +\\
    &+& \delta \left(- 4 n^{4} + 16 n^{3} - 24 n^{2} + 16 n - 4\right) + 2 n^{5} - 8 n^{4} + 12 n^{3} - 8 n^{2} + 2 n+\\
    &+& \sqrt{2\cdot (n-1)^{2} + 2 \left(n - 1\right)^{2}} \cdot(4 \delta^{3} - 3 \delta^{2} n - 2 \delta \left(n - 1\right)^{2} + n \left(n - 1\right)^{2})\\
    &=& 6 \delta^{5} - 4 \delta^{4} n + \delta^{3} \left(6 n^{2} - 4 n - 2\right) + \delta^{2} \left(- 6 n^{3} + 6 n^{2}\right) +\\
    &+& \delta \left(- 4 n^{4} + 12 n^{3} - 12 n^{2} + 4 n\right) + 2 n^{5} - 6 n^{4} + 6 n^{3} - 2 n^{2}.
\end{eqnarray*}

Similarly, as we did for $T(\delta)$, we can derive that $T_l(c_0n) < 0$, $T_u(c_0n) < 0$, $T_l(c_0n-0.5) > 0$, and $T_u(c_0n-0.5) > 0$ for $n > 2$. This implies the existence of zeros of the polynomials $T_l(\delta)$ and $T_u(\delta)$ in the interval $(c_0n-0.5, c_0n)$. Now, we aim to prove that these are the only zeros of the polynomials $T_l(\delta)$ and $T_u(\delta)$ within the interval $[1,n-1]$. 

The polynomials $T_l(c_0n)$ and $T_u(c_0n)$ are represented numerically as follows
\begin{eqnarray*}
    T_l(c_0n) &=& n^{4} \left(-1.144 + 0.065 \sqrt{2}\right) + n^{3} \left(2.731 - 0.605 \sqrt{2}\right) + n^{2} \left(-2.159 + 0.81 \sqrt{2}\right) + n \left(0.54 - 0.27 \sqrt{2}\right)\\
    T_u(c_0n) &=& - 1.014 n^{4} + 1.522 n^{3} - 0.54 n^{2}.
\end{eqnarray*}

For $n>2$ it holds that $n^{4} (-1.144 + 0.065 \sqrt{2}) + n^{3} (2.731 - 0.605 \sqrt{2}) < 0$ and $n^{2}(-2.159 + 0.81 \sqrt{2})\\ + n (0.54 - 0.27 \sqrt{2}) < 0$. Therefore, it holds that $T_l(c_0n) < 0$, for $n > 2$. On the other hand, by analyzing the quadratic equation $- 1.014 n^{2} + 1.522 n - 0.54$,  it is easy to see that $T_u(c_0n) < 0$, for $n>2$.

The polynomials $T_l(c_0n-0.5)$ and $T_u(c_0n-0.5)$ can be numerically represented as follows
\begin{eqnarray*}
    T_l(c_0n-0.5) &=& n^{4} \left(0.065 \sqrt{2} + 1.97\right) + n^{3} \left(-7.178 + 0.691 \sqrt{2}\right) + n^{2} \left(9.527 - 2.141 \sqrt{2}\right) + \\
    &+& n \left(-5.383 + 1.885 \sqrt{2}\right) - 0.5 \sqrt{2} + 1.0625\\
    T_u(c_0n-0.5) &=& 2.099 n^{4} - 5.796 n^{3} + 5.245 n^{2} - 1.613 n + 0.0625.
\end{eqnarray*}

For $n \geq 4$ it holds that $n^{4} (0.065 \sqrt{2} + 1.97) + n^{3} (-7.178 + 0.691 \sqrt{2}) > 0$ and $n^{2} (9.527 - 2.141 \sqrt{2}) \\+ n (-5.383 + 1.885 \sqrt{2}) - 0.5 \sqrt{2} + 1.0625 > 0$. It can be calculated for $n = 3$ that $T_l(c_0n-0.5) > 0$. Therefore, it holds that $T_l(c_0n-0.5) > 0$ for $n > 2$. On the other hand, we can analyze the summands of the polynomial $T_u(c_0n-0.5)$. For $n \geq 3$ it holds that $2.099 n^{4} - 5.796 n^{3} > 0$ and $5.245 n^{2} - 1.613 n > 0$, which implies that $T_u(c_0n-0.5) > 0$ for $n > 2$.

\bigskip

Now, we proceed with the analysis of the polynomials $T_l(\delta)$ and $T_u(\delta)$.
The first derivatives of $T_l$ and $T_u$ are equal to
\begin{eqnarray*}
    T_l'(\delta) &=& 30 \delta^{4} - 16 \delta^{3} n + \delta^{2} \left(18 n^{2} - 36 n + 12 \sqrt{2} n - 12 \sqrt{2} + 18\right) +\\
    &+& \delta \left(- 12 n^{3} - 6 \sqrt{2} n^{2} + 24 n^{2} - 12 n + 6 \sqrt{2} n\right) -\\
    &-& 4 n^{4} - 2 \sqrt{2} n^{3} + 16 n^{3} - 24 n^{2} + 6 \sqrt{2} n^{2} - 6 \sqrt{2} n + 16 n - 4 + 2 \sqrt{2}\\
    T_u'(\delta) &=& 30 \delta^{4} - 16 \delta^{3} n + \delta^{2} \left(18 n^{2} - 12 n - 6\right) + \delta \left(- 12 n^{3} + 12 n^{2}\right) - 4 n^{4} + 12 n^{3} - 12 n^{2} + 4 n.
\end{eqnarray*}

The number of sign changes between consecutive coefficients in both the polynomials $T_l'(\delta)$ and $T_u'(\delta)$ is equal to $3$, for $n>2$. According to Descartes' rule of signs, we can conclude that there are either one or three positive real roots of the polynomials $T'_l$ and $T'_u$. Assuming there are three real positive roots for these polynomials implies that all four roots are real numbers. By applying Gauss–Lucas theorem to polynomials $T'_l$ and $T'_u$ with real roots, it follows that their derivatives have three real roots. Therefore, we will examine the second derivatives of polynomials $T_l$ and $T_u$, which are given by

\begin{eqnarray*}
    T_l''(\delta) &=& 120 \delta^{3} - 48 \delta^{2} n + \delta \left(36 n^{2} - 72 n + 24 \sqrt{2} n - 24 \sqrt{2} + 36\right) - 12 n^{3} - 6 \sqrt{2} n^{2} + 24 n^{2} - 12 n + 6 \sqrt{2} n\\
    T_u''(\delta) &=& 120 \delta^{3} - 48 \delta^{2} n + \delta \left(36 n^{2} - 24 n - 12\right) - 12 n^{3} + 12 n^{2}.
\end{eqnarray*}

Upon analyzing the discriminants $D_l$ and $D_u$ of these cubic polynomials, given by

\begin{eqnarray*}
    D_l &=& - 35914752 n^{6} - 47195136 \sqrt{2} n^{5} + 177831936 n^{5} - 445699584 n^{4} +\\
    &+& 225856512 \sqrt{2} n^{4} - 460422144 \sqrt{2} n^{3} + 710581248 n^{3} - 697932288 n^{2} +\\
    &+& 487461888 \sqrt{2} n^{2} -  263761920 \sqrt{2} n + 373248000 n -82114560 + 58060800 \sqrt{2}\\
    D_u &=& - 35914752 n^{6} + 83441664 n^{5} - 49185792 n^{4} - 6967296 n^{3}+\\
    &+& 2820096 n^{2} + 4976640 n + 829440.
\end{eqnarray*}
we deduce that they are negative for $n>2$.
Ineed, we notice that for $n \geq 4$ the following inequalities hold $- 35914752 n^{6} - 47195136 \sqrt{2} n^{5} + 177831936 n^{5} < 0$, $- 445699584 n^{4} + 225856512 \sqrt{2} n^{4} -460422144 \sqrt{2} n^{3} + 710581248 n^{3} < 0$, $- 697932288 n^{2} + 487461888 \sqrt{2} n^{2} -263761920 \sqrt{2} n + 373248000 n < 0$, and $ -82114560 + 58060800 \sqrt{2} < 0$. For $n = 3$, by the direct computation, we see that $D_l < 0$, and hence $n \geq 3$. On the other hand, for $n \geq 4$ it holds that $- 35914752 n^{6} + 83441664 n^{5} < 0$, $-2820096n^{4} + 2820096n^2 < 0$, $-4976640n^4 + 4976640n < 0$, and $-829440n^4+829440 < 0$. Using the fact that $-2820096n^{4}-4976640n^4 -829440n^4 > -49185792n^4$, by summing the corresponding sides of these inequalities, we can conclude that $- 35914752 n^{6} + 83441664 n^{5}-2820096n^{4} + 2820096n^2-4976640n^4 + 4976640n-829440n^4+829440 < 0$ for $n > 3$, and therefore $D_u < 0$ for $n > 3$.  For $n=3$, by the direct computation, we obtain that $D_u < 0$. 
For $n>2$, both $D_l$ and $D_u$ are negative numbers, and  as the coefficients of $T_l''(\delta)$ and $T_u''(\delta)$ are real, it follows that there exist two complex non-real roots. This contradicts the assumption that the polynomials $T_l'(\delta)$ and $T_u'(\delta)$ each have four real roots, thereby implying that these polynomials possess precisely one positive real root. 

Furthermore, since $T_l'(0) < 0$ and $T_u'(0) < 0$ for $n > 2$, indicating that the polynomials $T_l$ and $T_u$ initially decrease from the point $\delta=0$ until the root of their derivatives, then increase across the remainder of the domain, it follows that the sole stationary points of $T_l$ and $T_u$ are local minima.

\medskip

\medskip

Suppose the only stationary point $x_0$ of the polynomial $T_l$ lies within the interval $(0, c_0n-0.5)$. Since $x_0$ is a local minimum, this implies that the polynomial $T_l$ increases within the interval $(c_0n-0.5, n-1)$. Consequently, $T_l(c_0n-0.5) < T_l(c_0n)$, leading to a contradiction, as we have previously established that $T_l(c_0n-0.5) < 0$ and $T_l(c_0n) > 0$. Since the local minimum $x_0$ belongs to the interval $(c_0n-0.5,n-1]$, we deduce that the polynomial $T_l(\delta)$ decreases within the interval $(0, c_0n-0.5)$. Thus, it follows that $T(\delta) \geq T_l(\delta) \geq T_l(c_0n-0.5) > 0$, for $\delta \in (0, c_0n-0.5)$, implying that there are no real zeros of the function $T(\delta)$ within this interval.

Moreover, by computation, we find that $T_u(n-1) = - 8 n^{4} + 28 n^{3} - 36 n^{2} + 20 n - 4$, which is negative for $n > 2$. The derivative of the polynomial $T_u(\delta)$ has only one positive real root, representing the local minimum. Therefore, we conclude that for $\delta \in (c_0n, n-1]$, the maximum occurs at the endpoints of the interval. Since it holds that $0 > \max{T_u(n-1), T_u(c_0n)} \geq T_u(\delta) \geq T(\delta)$ for $\delta \in (c_0n, n-1]$, we deduce that there are no zeros of the function $T(\delta)$ within this interval.

The interval $(c_0n-0.5, c_0n)$ encompasses all the zeros of the function $T(\delta)$ within the domain $\delta \in (1, n)$. Since vertex degrees are integer values, the only potential candidates for $\delta$ that achieve the maximum are $\lfloor c_0n \rfloor -1$, $\lfloor c_0n \rfloor$, and $\lceil c_0n \rceil$.

Let $k$ be one of these three values that attain the maximum. For $n > 2$, it can be calculated that $SO_5(M_{n, k}) \geq SO_5(S_n)$, where $S_n$ is a star graph of order $n$, with the equality holding only for $n = 3$ and $n = 4$.


\qed

\bigskip

Suppose that a graph $G$ attains the maximum of the function $SO_5$ and have three distinct values in its degree sequence, with minimum degree equal to $\delta$. We can observe that 
$$
SO_5(G) = 2\pi\sum_{(i,j) \in E(G)}f(d(i),d(j))\leq 2\pi\sum_{1\leq i<j\leq n-1 }f(d(i),d(j))\leq 2\pi\max\left(\sum_{1\leq i<j\leq n-1 }f(x_i,x_j)\right),
$$
where $\delta \leq x_i\leq n-1$.

The partial derivatives of $F(x_1,\ldots,x_n)=2\pi\sum_{1\leq i<j\leq n-1 }\frac{\mid x_i^2 - x_j^2 \mid}{\sqrt{2} + 2\sqrt{x_i^2 + x_j^2}}$ are equal to 

    \begin{equation*}
        \frac{\partial F}{\partial x_i} = 2\pi \cdot \sum_{ij \in E(G)} \frac{\frac{\mid x_i^2 - x_j^2 \mid}{x_i^2 - x_j^2} \cdot 2x_i (\sqrt{2} + 2\sqrt{x_i^2 + x_j^2}) +  \frac{\mid x_i^2 - x_j^2 \mid\cdot2x_i}{\sqrt{x_i^2+x_j^2}}}{2 + 4(x_i^2 + x_j^2) + 4\sqrt{2x_i^2 + 2x_j^2}}.
    \end{equation*}

It is evident that the only stationary points are $(x_1,\ldots,x_n)$ where $x_1=\cdots=x_n$, and the value of function $F$ is zero, indicating that these stationary points correspond to local (and global) minima. On the other hand, to analyze the maximum of the function, we only need to consider values $x_i$ that are boundary points of the domain.  Considering graphs with three distinct values in their degree sequence, the coordinates $x_i$ of the boundary points $(x_1,\ldots,x_n)$ can assume the values $\delta$, $n-2$, or $n-1$.
Since in some of these points the function $F$ achieves the maximum, it does not mean that there exists a graph $G$ such that the degrees of its vertices belong to the set $\{\delta,n-2,n-1\}$. However, the fact that the function $F$ achieves its maximum at some of these points does not imply the existence of a graph $G$ where the degrees of its vertices belong to the set $\{\delta,n-2,n-1\}$. Therefore, for the remainder of this section, we will investigate graphs with the degree sequence $(\underbrace{n-1,\ldots,n-1}_{l}, \underbrace{n-2,\ldots,n-2}_{k}, \underbrace{\delta,\ldots,\delta}_{n-k-l})$ and illustrate the complexity of determining the maximum value of $SO_5$ within this class of graphs. Furthermore, based on computer search tests, we conclude that extremal graphs, concerning the maximum $SO_5$ within the class of graphs with degree sequences featuring three distinct values, should be sought within the subset where degree sequence values belong to the set $\{\delta,n-2,n-1\}$.

First, we can show, using the same method as employed earlier in this section, that the vertices with degrees $n-1$ form a clique of size $l$ as a subgraph within $G$, while vertices with degrees $\delta$ constitute an independent set. The vertices with degree $n-2$ are connected to all vertices in the graph except one, which has either degree $n-2$ or degree $\delta$. Let us assume that the number of vertices with degree $n-2$, which are not connected to a vertex of degree $\delta$, is denoted as $s$, where $0\leq s\leq k$. Let $\delta_1$ represent the number of neighbors among these $s$ vertices for an arbitrary vertex of degree $\delta$. 
The number of edges connecting vertices with degree $\delta$ and $s$ vertices with degree $n-2$ is equal to $(n-k-l)\delta_1$.
On the other hand, given that each of these $s$ vertices has $n-k-l-1$ neighbors among the $n-k-l$ vertices of degree $\delta$, the total number of edges connecting these $s$ vertices to $n-k-l$ vertices of degree $\delta$ is $s(n-k-l-1)$, implying that 
$$
(n-k-l)\delta_1=s(n-k-l-1).
$$

Based on this equation, we derive the following cases: either $n-k-l$ divides $s$ and $n-k-l-1$ divides $\delta_1$, or $\delta_1 = s = 0$.
In the first case, we can conclude that $s=n-k-l$ and $\delta_1=n-k-l-1$, as we can establish a bijection between an arbitrary vertex belonging to the set of $s$ vertices with degree $n-2$ and a vertex with which it is not connected from the set of $n-k-l$ vertices with degree $\delta$.
After substituting $s$ and $\delta_1$ into the last equation, where $\delta = l + (k - s) + \delta_1$, we finally obtain $\delta = k + l - 1$.
Given that the degree sequence of $G$ consists of only three distinct values, we find that
\begin{eqnarray}
    \label{SO5_for three distinct}
    SO_5(G)=m_{n-1,n-2}f(n-1,n-2)+m_{n-1,\delta}f(n-1,\delta)+m_{n-2,\delta}f(n-2,\delta).
\end{eqnarray}
Substituting $k=\delta-l+1$ into the equation for $SO5$, we obtain
$$
SO_5(G_1)=(\delta-l+1)lf(n-1,n-2)+(n-\delta-1)lf(n-1,\delta)+(n-\delta-1)(\delta-l)f(n-2,\delta).
$$
In the second case, when $\delta_1=s=0$, it is clear that $\delta=k+l$, which implies $k=\delta-l$.
Substituting $k$ into the equation for (\ref{SO5_for three distinct}), we get
$$
SO_5(G_2)=(\delta-l)lf(n-1,n-2)+(n-\delta)lf(n-1,\delta)+(n-\delta)(\delta-l)f(n-2,\delta).
$$
By subtracting  $SO_5(G_2)$ from $SO_5(G_1)$, we get that 
$$
SO_5(G_1)-SO_5(G_2)=lf(n-1,n-2)-lf(n-1,\delta)-(\delta-l)f(n-2,\delta)<0,
$$
which leads to the conclusion that the maximal value of $SO_5$ should be sought within the class of graphs satisfying the conditions of the second case.

We can observe that $SO_5(G_2)$ represents a two-variable function, denoted by $G(\delta,l)$. The partial derivatives of  $G(\delta,l)$ are
\begin{eqnarray*}
 \frac{\partial G}{\partial \delta}&=&lf(n-1,n-2)-l\frac{\partial (\delta\cdot f(n-1,\delta))}{\partial \delta}-l\frac{\partial ((n-\delta)\cdot f(n-2,\delta))}{\partial \delta}+\frac{\partial (\delta(n-\delta)\cdot f(n-2,\delta))}{\partial \delta}\\
 \frac{\partial G}{\partial l}&=&-2lf(n-1,n-2)+(n-\delta)f(n-1,\delta)-(n-\delta)f(n-2,\delta).
\end{eqnarray*}

It is evident that determining the stationary points of $G(\delta,l)$ is a much more challenging problem than finding the stationary points of the function $F(\delta)=\delta(n-\delta)\cdot f(n-1,\delta)$, which was a key step in the previous theorem. The problem becomes significantly more complex when viewed as a constrained optimization problem with the constraints $2\leq \delta\leq n-3$ and $1\leq l\leq \delta-2$. Based on computer experiments, it is observed that the maximum value of the function $G(\delta,l)$ under these constraints, where $\delta$ and $l$ are real numbers, is lower than the maximum value of $F(\delta)$, where $2\leq \delta\leq n-2$ is an integer, as obtained in the preceding theorem. Therefore, we posit that the maximum of $F(\delta)$ corresponds to the maximum of $SO_5$ within the entire class of connected graphs, and we defer further investigation to future studies.


\section{Concluding remarks}

In this paper, we establish the maximum values of $SO_5$ and $SO_6$ within the set of molecular trees with a specified number of vertices. Additionally, we determine the maximum value of $SO_5$ within the set of graphs obtained by applying the join operation to two specific graphs of a given order. The proofs presented in this paper are generally grounded in the connection between number theory, polynomial theory, and multivariable function analysis, spanning numerous distinct cases.
Fully classifying the set of connected graphs with the maximum $SO_5$ would be considerably more demanding,  involving the examination of more complex multivariable functions and potentially a significantly larger number of cases. It is worth noting that in the analysis of univariate functions, the problem is reduced to the analysis of fifth-degree polynomials, which generally do not have solutions in all cases. The task of finding the maximum $SO_6$ becomes even more intricate, even though in the scenario of addressing univariate functions, it can be reduced to the analysis of tenth-degree polynomials, which we defer for future research.

\end{document}